\documentclass[english,svgnames,dvipsnames,table]{svjour3}
\usepackage[T1]{fontenc}
\usepackage[latin1]{inputenc}
\usepackage{geometry}
\usepackage{color}
\usepackage{array}
\usepackage{float}
\usepackage{url}
\usepackage{bm}
\usepackage{multirow}
\usepackage{amsmath}
\usepackage{amssymb}
\usepackage{graphicx}

\makeatletter

\providecommand{\tabularnewline}{\\}

\newcommand{\vl}{\vline width0pt height 12pt depth 6pt}

\AtBeginDocument{
  
}

\makeatother

\usepackage{babel}
\begin{document}
\title{Counting the uncounted : estimating the unaccounted COVID-19 infections
in India}
\author{Debashis Saikia, Kalpana Bora, and Madhurjya P. Bora}
\institute{Physics Department, Gauhati University, Guwahati 781014, India}
\maketitle
\begin{abstract}
Undetected infectious populations have played a major role in the
COVID-19 outbreak across the globe and estimation of this undetected
class is a major concern in understanding the actual size of the COVID-19
infections. Due to the asymptomatic nature of some infections, many
cases have gone undetected. Also, despite carrying COVID-19 symptoms,
most of the infected population kept the infections hidden and stayed
unreported, especially in a country like India. Based on these factors,
we have added an undetected compartment to the already developed SEIR
model \cite{OurModel} to estimate these uncounted infections. In
this article, we have applied Physics Informed Neural Network (PINN)
to estimate the undetected infectious populations in the $20$ worst-affected
Indian states as well as India as a whole. The analysis has been carried
out for the first as well as second surge of COVID-19 infections in
India. A ratio of the active undetected infectious to the active detected
infectious population is calculated through the PINN analysis which
gives a picture of the real size of the pandemic in India. The rate
at which symptomatic infectious population goes undetected and are
never reported is also estimated using the PINN method. Toward the
end, an artificial neural network (ANN) based forecasting scenario
of the pandemic in India is presented. The prediction is found to
be reliable as the training of the neural network has been carried
out using the unique features, obtained from the state-wide analysis
of the newly proposed model as well as from the PINN analysis.
\end{abstract}

\section{1. Introduction}

The impact of the COVID-19 pandemic has been devastating and almost
all the countries have experienced this new disease, primarily in
different phases at different times. Researchers around the globe
are putting their constant effort to understand the dynamics of the
transmission of the disease. We have also seen extraordinary efforts
to roll out the COVID-19 vaccines within a very short time, which
could save millions of lives. Many COVID-19-specific mathematical
models have been developed over the course of the pandemic which can
help us understand the way the COVID-19 disease has spread \cite{model1,model2,model3,model4,model5,model6}.
These results can be really helpful for policymakers to make strategies
to suppress future similar disease spread. Artificial neural networks
(ANN), which have a wide ranging applicability due to their universality,
have already been used to predict disease outbreaks \cite{pred1,pred2,pred3}
and detection \cite{xray1,xray2,xray3}, including COVID-19. Many
studies also have found the method of `fractional derivative' to be
an efficient method in studying disease modelling \cite{ghanbari-1,ghanbari-2,ghanbari-3,ghanbari-4,ghanbari-5,ghanbari-6,ghanbari-7,ghanbari-8}.

Most of the countries in the world have experienced the COVID-19 disease
in distinct phases. So far as the pandemic in India is concerned,
it has experienced the pandemic in two distinct phases, the first
phase effectively started in March 2020 and the second surge of infections
started at the end of February 2021. In India, the second phase of
the pandemic was observed to be more contagious and lethal compared
to the first phase. The rise of the spike in COVID-19 cases in the
second phase is now believed to be mainly due to the $\delta$-variant
of the SARS-CoV-2 virus. New strains of the virus such as double mutant
variant (B.1.617) and triple mutant variant (B.1.618) were seen to
be more vulnerable when the population density is high \cite{mutant1}.
The $\delta$-variant of the SARS-CoV-2 virus (B.1.617.2), which was
detected in India towards the end of 2020 was highly infectious and
soon became a key reason behind the sudden surge of infections during
the second phase of the pandemic \cite{mutant2}. Unlike the first
phase of the pandemic, more younger people were affected in the second
phase. Till mid-December 2021, India recorded $\sim35$ million confirmed
COVID-19 cases with $476$ thousand deaths while the tally of confirmed
cases has crossed $270$ million globally with more than $5.3$ million
deaths \cite{John}.

India, being the most populous country in the world now, presents
certain unique scenarios as far as the COVID-19 pandemic is concerned.
Considering the size of the population in a country like India, while
it is of great interest to correctly model the progress of a pandemic
to correctly understand the overall global scenario, we at the same
time have to deal with some uncertainties with reference to detailed
and correct patient data, which are available through verifiable sources.
In the case of the COVID-19 pandemic, one primary concern in this
regard was the actual number of infected people against what was reported
or available officially. There are many reasons why a discrepancy
may exist between these two, many of which can not probably be controlled
due to widely varying social, economic, and geographical conditions
that exist in the country. Our aim of this work is to see what might
be the actual size of the infected people, many of whom might not
be counted officially, in a pandemic which is still evolving. This
may help us understand situations in other countries with similar
social and economic conditions and help us understand the future pandemic
progression. In this work, we have used our previously and successfully
developed SEIR (susceptible-exposed-infectious-removed) model \cite{OurModel}
and have come up with a new SEIUR (susceptible-exposed-infectious-undetected-removed)
which is being solved with the help of artificial neural network (ANN).
We have applied a relatively new technique known as the Physics Informed
Neural Network (PINN) \cite{lagaris} to estimate the undetected infectious
population and also have used this technique to determine the unknown
parameters of the model known as the so-called parameter discovery.
Based on our results from the PINN model, we have also proposed a
PINN-ANN-based prediction scenario, which might be used during similar
situations in the future. We note that artificial neural network (ANN)
is rapidly becoming a general purpose tool which has found wide-ranging
applications in extremely diverse situations. In this work, we have
applied the PINN analysis for the so-called \textquoteleft parameter
discovery\textquoteright{} of our newly developed SEIUR model, which
itself is based on our very successful SEIR model applied to the COVID-19
outbreak in India. Though, individually all the methods have already
been tested and developed, this is the first time that PINN is being
applied for parameter discovery of an epidemiological model.

In section 2, we discuss the development of the SEIUR model and the
basic reproduction number is calculated. In section 3, modelling using
the neural network has been carried out and the model has been solved
using the PINN tool for India as well as for 20 Indian states. In
section 4, we present a possible forecast scenario of the COVID-19
outbreak in India based on the first and second phases of the pandemic.
In section 5, we conclude.

\section{2. The epidemiological model}

\subsection{{\bf 2.1 The SEIR model}}

In a disease like COVID-19, the role of the `exposed' population becomes
very important in mathematical modelling. In our earlier work, we
proposed an SEIR model \cite{OurModel} (will be referred to as the
SEIR model, hereafter), designed specifically to deal with the unavailability
of detailed patient data, which has been successfully applied to predict
the progress of the pandemic in the first phase (from March 2020 to
February 2021), well beyond the available data \cite{OurModel}. The
equations of this model are given by
\begin{eqnarray}
\frac{dS}{dt} & = & \lambda-(\beta_{t}+\rho_{t})\frac{IS}{N}-\delta S,\label{eq:seir1}\\
\frac{dE}{dt} & = & \beta_{t}\frac{IS}{N}-(\nu_{t}+\delta)E,\\
\frac{dI}{dt} & = & \nu_{t}E-(\gamma+\delta)I+\rho_{t}\frac{IS}{N},\\
\frac{dR}{dt} & = & \gamma I-\delta R,\label{eq:seir4}
\end{eqnarray}
where, $S,E,I$, and $R$ are the susceptible, exposed, infectious
and removed population, respectively, at any point of time. The removed
compartment includes the population who have recovered from the disease
or have died from the disease. The variables satisfy the condition
\begin{equation}
S(t)+E(t)+I(t)+R(t)=N={\rm Const.}
\end{equation}
In this model, we have taken care of the detected asymptomatic population
with the \emph{zero} incubation period \cite{OurModel} and linked
them directly to the infectious group without taking an extra compartment
for the asymptomatic infectious population. Here, the asymptomatic
populations are those who do not develop any symptoms but are infectious
and can be detected only via contact tracing or random testing. The
two transmission rates $\beta_{t}$ (disease transmission rate) and
$\rho_{t}$ (asymptomatic transmission rate) are assumed to be time-dependent
piecewise functions. Piecewise functions are used to capture the changes
in an ongoing pandemic due to varying conditions. The incubation period
is expressed through a time-dependent function, which starts at a
very large number approaching a constant value of $14$ days. The
parameters $\delta$ and $\gamma^{-1}$ are the natural death rate
and recovery period, respectively.

As mentioned before, one primary concern is to estimate the so-called
\emph{undetected} population, which can be sizable and never contribute
to the officially recorded number of patients. It is very important
to distinguish the \emph{undetected }population\emph{ }from the \emph{detected
asymptomatic }population. While the former is the asymptomatic populations
who are detected via random testing and being recorded officially
as infected, the latter is the asymptomatic population that has gone
undetected. In both cases, they do not develop any symptoms but are
infectious.
\begin{center}
\begin{figure}[t]
\begin{centering}
\includegraphics[width=0.6\textwidth]{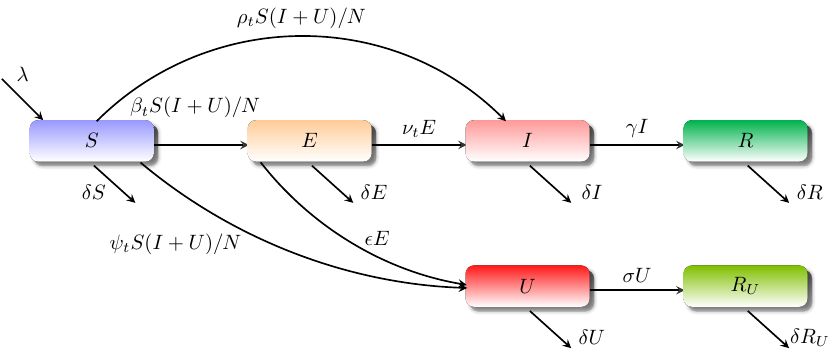}
\par\end{centering}
\caption{\label{fig:Compartmental-diagram-of}Compartmental diagram of the
newly proposed SEIUR model.}
\end{figure}
\par\end{center}

\subsection{{\bf 2.2 The SEIUR model}}

\subsection{\bf{2.2.1 Assumptions}}

So far, India has seen three distinct phases of the COVID-19 pandemic
--- the first phase which lasted from January 2020 to February 2021
and the second phase which started in late February, 2021 and peaked
in around mid-April, 2021 and subsided with a very prolonged and tapered
tail till the end of December, 2021 after which the third phase had
started. The second phase of the pandemic was believed to have been
driven by the $\delta$-variant of the SARS-COV-2 virus, driving a
huge surge of infections within a very short period of time \cite{mutant2}.
In this paper, we have only considered the first and second phases
of the pandemic.

Despite the visible difference between the two phases of the pandemic,
the overall ambient socio-economic conditions during both phases of
the pandemic remained almost the same. As the vaccination of the population
did not start till late March 2021, with considerable accuracy it
can be assumed that in both phases of the pandemic, vaccination had
only a little role to play. With these observations, we can safely
assume that the primary nature of the pandemic in both phases remained
the same with only a larger infection rate during the second phase
due to the highly infectious $\delta$-variant of the virus. As such,
our compartmental model should remain the same with a difference \emph{only}
in the infection rates. We further assume that all the assumptions
that have been made in our previously described model (which was applied
only to the first phase of the pandemic) \cite{OurModel} should remain
valid during both phases of the pandemic.

With these assumptions in place, we now propose a modified SEIUR (Susceptible-Exposed-Infectious-Undetected-Removed)
compartmental model with an attempt to estimate the undetected population.
Here, the undetected population comprises both asymptomatic undetected
population and those who intentionally hide their disease and are
never reported as infected individuals. The undetected group of the
population plays an important role in the transmission of the disease
in a particular locality. There are some social factors also, which
have fuelled the pandemic at different times. To incorporate the significance
of these two classes of the undetected population, an extra compartment
for the undetected population is added to our previously reported
SEIR model. An \emph{undetected-removed} compartment is also added
to the new model for the sake of conservation of the total population.
In the new model, we have introduced another infective class $U$
to represent the undetected population. There is always a probability
that the interaction between $U$ and $S$ will lead to a more infectious
population, who is further reported as detected infectious or may
move to the undetected compartment. Similarly, the interaction between
$I$ and $S$ also results in an infectious population. The more the
interactions between $S$ and $I$, and $S$ and $U$ are, the more
the infectious count will be. The schematic model of this new SEIUR
model is shown in Fig.\ref{fig:Compartmental-diagram-of}.

The governing equations of the newly proposed SEIUR model are given
by
\begin{align}
\frac{dS}{dt} & =\lambda-(\ensuremath{\psi_{t}}+\beta_{t}+\rho_{t})\frac{(I+U)S}{N}-\delta S,\label{eq:seiur1}\\
\frac{dE}{dt} & =\beta_{t}\frac{(I+U)S}{N}-(\nu_{t}+\text{\ensuremath{\epsilon}+}\delta)E,\\
\frac{dU}{dt} & =\psi_{t}\frac{(I+U)S}{N}+\text{\ensuremath{\epsilon E}}-\text{(\ensuremath{\sigma}}+\delta)U\\
\frac{dI}{dt} & =\nu_{t}E-(\gamma+\delta)I+\rho_{t}\frac{(I+U)S}{N},\\
\frac{dR}{dt} & =\gamma I-\delta R,\label{eq:seiur5}\\
\frac{dR_{U}}{dt} & =\sigma U-\delta R_{U},\label{eq:seiur6}
\end{align}
where $U$ and $R_{U}$ represent the \emph{undetected} and \emph{undetected-removed}
population, respectively at any given time. The rest of the variables
are the same as stated above. Here we introduce a new transmission
rate -- `undetected transmission rate' denoted by $\psi_{t}$, which
is an important parameter as it determines how fast the pandemic will
rise or decline, while $\sigma^{-1}$ is the recovery rate for the
undetected population and the $\epsilon$ denotes the rate at which
symptomatic individuals \emph{hide} their disease (i.e.\ prevent
themselves from being recorded officially as infected). We can call
$\epsilon$ as the rate of \emph{symptomatic unreported} individuals.

\begin{figure}[t]
\begin{centering}
\includegraphics[width=0.5\textwidth]{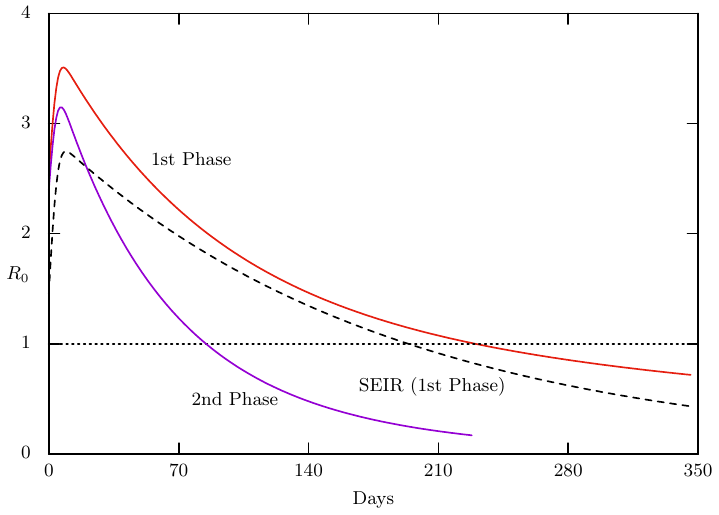}
\par\end{centering}
\caption{\label{fig:Variation-of-}Variation of $R_{0}$ with time for the
first and second phases with the new SEIUR model and for the first
phase with the previous SEIR model.}
\end{figure}

\subsection{\bf{2.2.2. Modelling the rates}}

Our gross assumption in formulating the new SEIUR model is that the
overall evolution of the COVID-19 pandemic in India is correctly modelled
by our SEIR model \cite{OurModel} thereby ascertaining the continuation
of the basic assumptions made earlier regarding the transmission rates.

We note that the transmission rates of the disease change with time
and so, the transmission rates are expressed in terms of time-dependent
piecewise functions expressed in certain forms. The transmission rates
$\beta_{t}$ and $\rho_{t}$ are the same as in the SEIR model. The
time-dependent incubation period $\nu_{t}$ is also kept unchanged.
Here we introduce the new transmission rate $\psi_{t}$ in terms of
the disease transmission rate $\beta_{t}$ for the undetected population
\begin{eqnarray}
\beta_{t} & = & \alpha_{t}e^{-\mu_{t}t},\label{eq:beta-1}\\
\psi_{t} & = & \beta_{t}e^{\zeta_{t}t}\\
\rho_{t} & = & \eta_{t}{\rm erf}(\rho_{0}t),\label{eq:rho-1}\\
\nu_{t} & = & \nu_{0}e^{-\varphi t}+\nu_{f},\label{eq:nu-1}
\end{eqnarray}
where $\mu_{t}=\zeta_{t}\kappa_{t}$ with $\kappa_{t}$ is a piecewise
function. We note that, from the analysis of the SEIR model \cite{OurModel},
the asymptomatic transmission rate $\eta_{t}$ was found to be $\sim15\%$
of the disease transmission rate $\beta_{t}$ for the first phase
of the pandemic. Therefore, in this new SEIUR model, we can safely
retain the same relation due to the same prevailing conditions during
both phases of the pandemic. The recovery periods $\sigma^{-1}$ and
$\gamma^{-1}$ are taken to be $21\,{\rm days}$. The natural death
rate $\delta$ is neglected with the assumption that the deaths caused
by COVID-19 during the pandemic period are very high compared to the
natural deaths. The new parameters introduced in the SEIUR model are
$\kappa_{t}$ and $\epsilon$.

\begin{table}[t]
\caption{\label{tab:Neural-network-parameters}Neural network parameters}
~\\

\centering{}{\small{}}%
\begin{tabular}{|l|c|}
\hline 
\multicolumn{1}{|c|}{\textbf{Parameters}} & \textbf{Value}\tabularnewline
\hline 
\hline 
Hidden layers\vl  & $8$\tabularnewline
\hline 
{\small{}Neurons in each layer}\vl  & $64$\tabularnewline
\hline 
Total neurons\vl  & $8\times64=512$\tabularnewline
\hline 
Activation function\vl  & $\tanh$\tabularnewline
\hline 
Optimization\vl  & Adam\tabularnewline
\hline 
\end{tabular}{\small\par}
\end{table}

\subsection{\bf{2.2.3. Basic reproduction number $\left(R_{0}\right)$}}

The basic reproduction number $R_{0}$ is a measure of the contagiousness
of an infectious disease that defines the average number of secondary
infections that stem out from a primary infection. Naturally, $R_{0}>1$
means an outbreak. For the SEIUR model, the basic reproduction number
can be calculated as
\begin{equation}
R_{0}=\frac{1}{(\sigma+\delta)}\left[\psi_{t}+\frac{\epsilon\beta_{t}}{(\nu_{t}+\epsilon+\delta)}\right]+\frac{1}{(\gamma+\delta)}\left[\rho_{t}+\frac{\nu_{t}\ensuremath{\beta_{t}}}{(\nu_{t}+\epsilon+\delta)}\right],
\end{equation}
which is plotted in Fig.\ref{fig:Variation-of-}, for both the previous
SEIR (first phase) and the new SEIUR (first and second phases) models.
As is seen from the figure, $R_{0}$ is larger during the first phase
of the pandemic compared to the second and both the SEIR and SEIUR
models closely agree with the behaviour during the first phase.

The days indicated in the figure start from the `zeroth' day, which
in the case of the first phase, indicates the day when the first COVID-19
cases were detected in India on 14 March 2020. We also note that,
the first case in India was detected on 30 January 2020 in the state
of Kerala. After that only a few cases were reported till mid-March
2020 and in our analysis, we have taken zeroth day as 14 March, 2020,
the first day when the number of active cases saw a significant jump.
The `zeroth' day for the second phase is taken as the day when the
number of daily infected cases rose for the first time after the initial
decline of the cases in the first phase, which is on the 24th of February
2021.

We note here the calculation of $R_{0}$ always require and underlying
numerical model, based on which the $R_{0}$ value can be determined
\cite{r0-new-1,r0-new-2}. Naturally, different models yield different
vales for $R_{0}$ \cite{r0-new-3}. The agreed value of $R_{0}$
is thus due the model, which can provide the most reliable pandemic
scenario based on modeling of the available data.

\begin{figure}[t]
\begin{centering}
\includegraphics[width=0.5\textwidth]{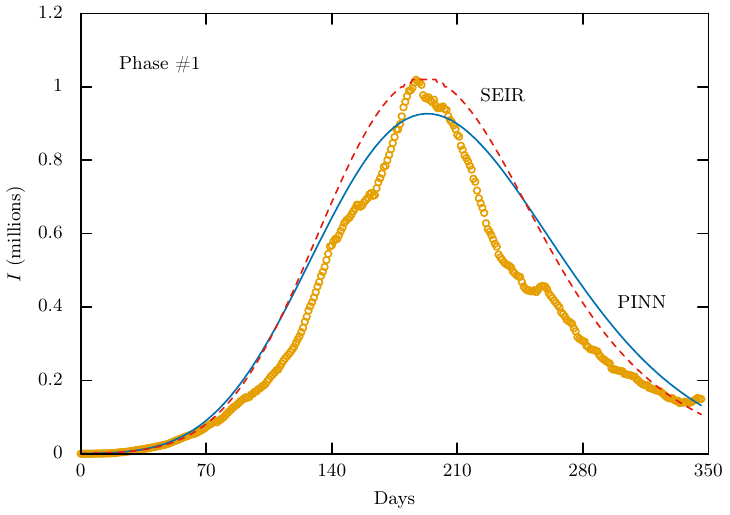}\hfill{}\includegraphics[width=0.5\textwidth]{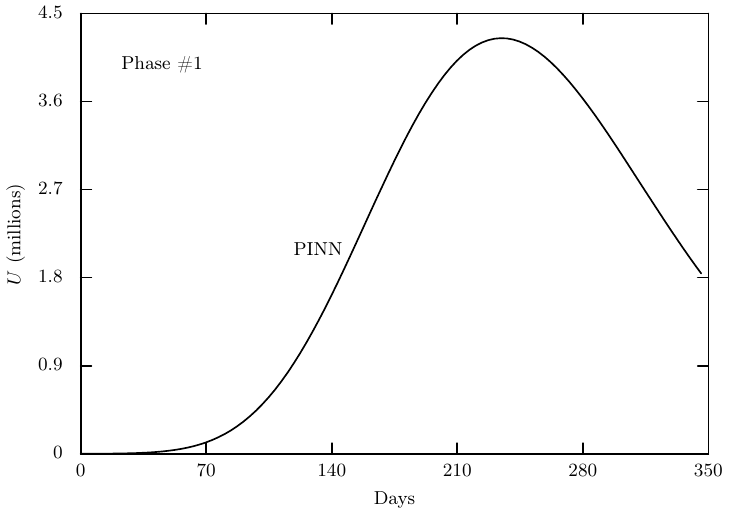}\\
\includegraphics[width=0.5\textwidth]{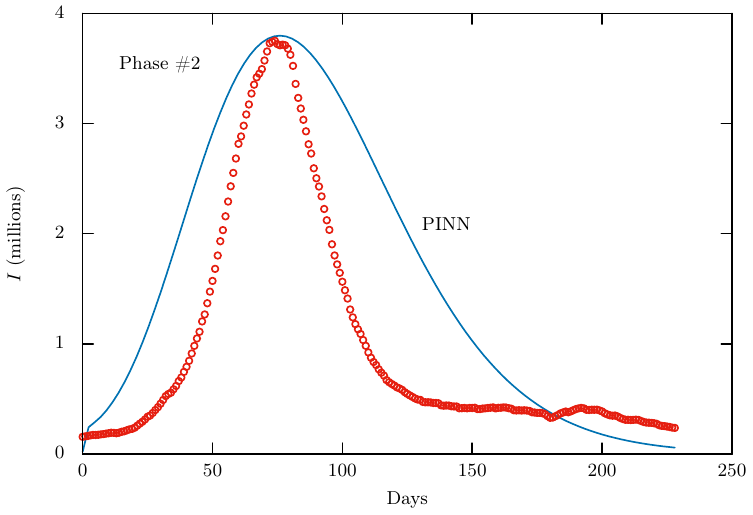}\hfill{}\includegraphics[width=0.5\textwidth]{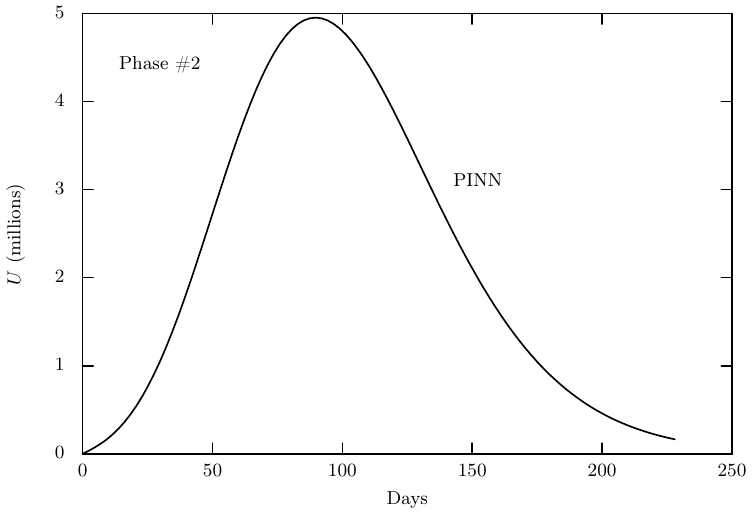}\\
\includegraphics[width=0.5\textwidth]{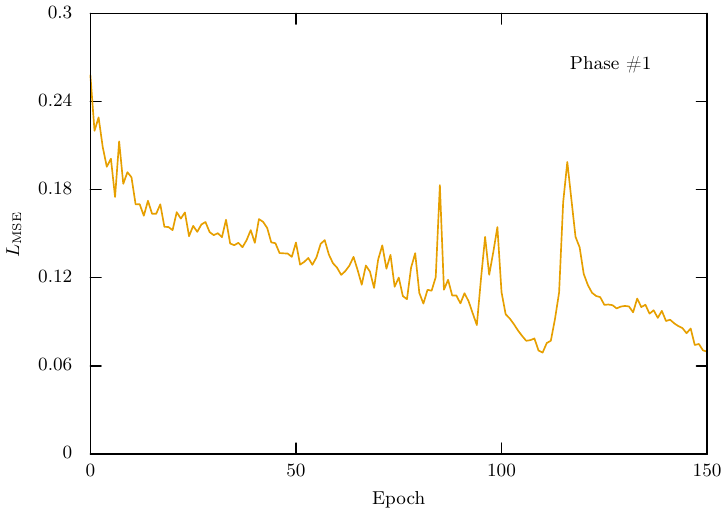}\hfill{}\includegraphics[width=0.5\textwidth]{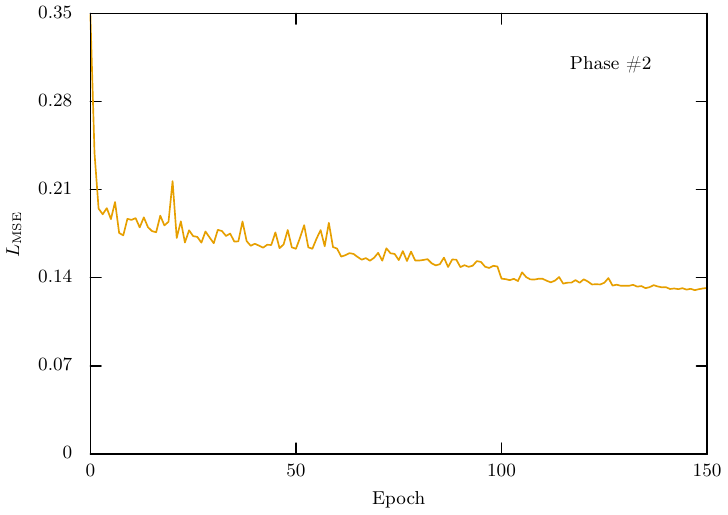}
\par\end{centering}
\caption{\label{fig:The-PINN-solutions}The PINN solutions of the new SEIUR
model applied to the first phase (Phase \#1) and the second phase
(Phase \#2) of the COVID-19 pandemic in India for the active infected
population $(I)$ and the active undetected population $(U)$. The
actual data points are indicated through the open circles `$\circ$'.
The bottom panel shows the model loss calculated in terms mean squared
error.}
\end{figure}

\section{3. Neural network modelling}

Of late, artificial neural network, commonly known as ANN has become
an indispensable part of the field of science and technology and has
made their way deep into almost all fields of science. Physics informed
neural network (PINN) \cite{lagaris} is relatively a new concept
that has emerged from the ANN and has gained importance due to its
universality of applications. PINN is an optimization method, through
which differential equations are treated as an optimization problem
with embedded initial or boundary conditions, which is then solved
using ANN. The name PINN, which originated from physics-related equations
\emph{does not }however limit its use \emph{only }to physics problems.
In this work, we have successfully applied the concept of PINN to
our SEIUR model. One important advantage of using the PINN approach
is that it can discover unknown parameters from the given dataset.
The unknown parameter $\epsilon$ in our model is in fact estimated
through this PINN approach. For a detailed analysis of the PINN approach,
the reader can see the work by \cite{lagaris,haghighat1,raissi}.

At this point, it is worthwhile to mention some of the existing methods
in epidemiological modelling, which can help the reader distinguish
the difference between the PINN approach and others. One class of
modelling is the so-called regression modelling, which as the name
suggests, obtains the pandemic properties through regression which
may include various approaches such as logistic regression, Bayesian
ridge regression, and Gaussian regression \cite{regression-1,regression-2,regression-3,regression-4,OurModel}.
These methods are useful when the pandemic datasets are very large
and have relatively less complex and interlinked parameters. In contrast
to this ANN-based methods can handle extremely complex interlinked
data, requiring large number of underlying differential equations
\cite{ann-1,ann-2,ann-3}. While various ANN-based methods can be
applied to many different stages of the modelling, the PINN approach
is used to optimise the unknown interlinked parameters which are used
in the underlying model.

\subsection{{\bf 3.1. The PINN setup}}

We now set up Eqs.(\ref{eq:seiur1}-\ref{eq:seiur5}) through a neural
network and assume that our field variables $\bm{\chi}=(S,E,I,U,R)$
can be approximated through a neural network ${\cal N}(t)$ 
\begin{equation}
\bm{\chi}(t)\equiv{\cal N}(t).
\end{equation}
The corresponding loss functions are
\begin{equation}
L_{i}(t)=\dot{\chi}_{i}(t)-f_{i}(\chi_{i},p_{j}),\quad(i,j)=1,2,\dots,5,
\end{equation}
where $f_{i}(\chi_{i},p_{j})$ are the right-hand sides of Eqs.(\ref{eq:seiur1}-\ref{eq:seiur5})
and $p_{j}=(\alpha_{t},\kappa_{t},\zeta_{t},\epsilon,\varphi)$ are
the parameters. The parameters $(\alpha_{t},\kappa_{t},\epsilon)$
are those which are to be determined through neural network optimization.
The initial conditions are further expressed in a set of another five
loss functions \cite{lagaris}
\begin{equation}
L_{k}(t)=[1-{\rm sign}(t)]\left[\dot{\chi}_{k}(t)-\chi_{0}\right],\quad k=1,2,\dots,5,
\end{equation}
where $\chi_{0}$ are the corresponding initial conditions and the
function ${\rm sign}(t)$ is defined as
\begin{equation}
\begin{array}{rclll}
{\rm sign}(t) & = & \pm1 &  & {\rm if}\,t\lessgtr0,\\
 &  & 0 &  & {\rm if}\,t=0.
\end{array}
\end{equation}
Note that the last Eq.(\ref{eq:seiur6}) does not need to be included
in the optimization problem as it \emph{does not} affect the rest
of the equations. The data sets which are to be used throughout the
optimization process were taken from the publicly sourced repository
at https://www.covid19india.org \cite{covid19}.

\textcolor{black}{}
\begin{table}[t]
\textcolor{black}{\caption{\label{tab:The-summary-of}The summary of PINN results for India}
~}\\

\centering{}\textcolor{black}{}%
\begin{tabular}{|c|c|c|c|c|c|}
\hline 
\textcolor{black}{Sl. No.\vl} & \textcolor{black}{\,\,Phase\,\,} & \textcolor{black}{$U_{{\rm peak}}$} & \textcolor{black}{$I_{{\rm peak}}$} & \textcolor{black}{$\epsilon$} & \multicolumn{1}{c|}{\textcolor{black}{$U_{{\rm peak}}/I_{{\rm peak}}$}}\tabularnewline
\hline 
\hline 
\textcolor{black}{1\vl} & \#1 & $\:4.2627\:$ & $\ 0.9312\ $ & $\ 0.0484\ $ & $4.5776$\tabularnewline
\hline 
\textcolor{black}{2\vl} & \textcolor{black}{\#2} & \textcolor{black}{$4.9605$} & \textcolor{black}{$3.8079$} & \textcolor{black}{$0.0493$} & \textcolor{black}{$1.3027$}\tabularnewline
\hline 
\end{tabular}
\end{table}

\begin{figure}[t]
\begin{centering}
\includegraphics[width=1\textwidth]{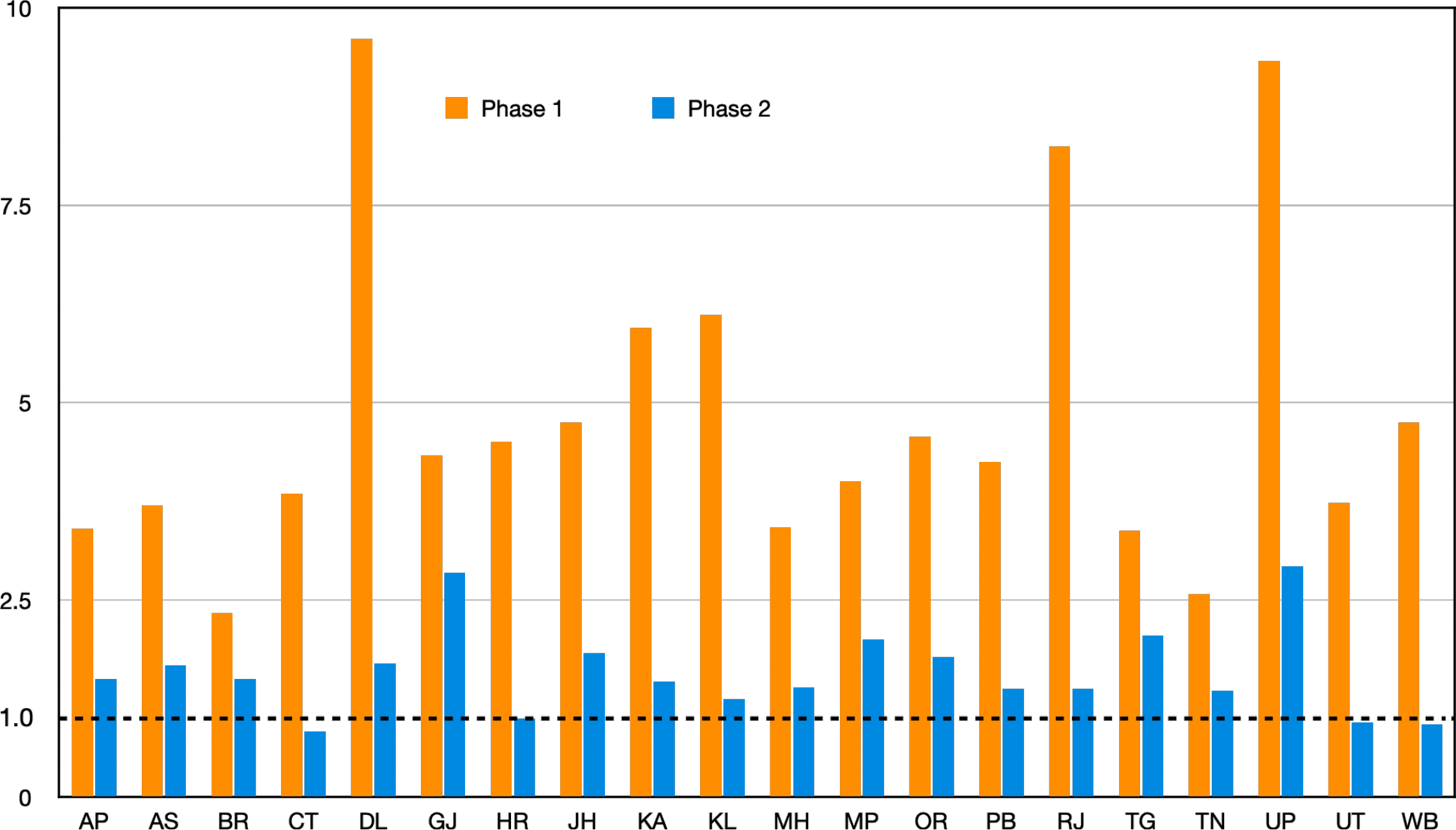}
\par\end{centering}
\caption{\label{fig:The-active-ratio}The active ratio $U/I$ - of the undetected
to the detected population for 20 states.}
\end{figure}

We have solved the system using a neural network consisting of $8$
hidden layers, each with $64$ neurons with kernel regularization.
The activation function used is `$\tanh$' and the optimization scheme
is Adam (see Table.\ref{tab:Neural-network-parameters}). Our working
interface is the SciANN package, which is basically a Tensorflow-Keras
wrapper \cite{haghighat1}.

\subsection{{\bf 3.2 PINN results}}

The PINN results for both phases of the pandemic in India are summarized
in Table.\ref{tab:The-summary-of} and the results are shown in Fig.\ref{fig:The-PINN-solutions}.
From the analysis, it is seen that the ratio of the active undetected
population to that of the active detected population was more during
the first phase of the pandemic compared to the second. In the first
phase, for every detected infection, there were almost $5$ persons
who went undetected. Interestingly, this scenario is found to be different
for the second phase of the pandemic. The count of undetected population
decreases at the second phase of the pandemic and we have just one
undetected individual for each detected infectious person.

A state-wise analysis has been done for the $20$ most affected states
of India up to October $10$, $2021$. This analysis is indeed important
as all the states are different from each other in terms of economy,
demography, and diversity in culture and lifestyle. A state-wise analysis
can also provide a picture of the response of Indian states towards
the COVID-19 outbreak. The PINN results of the COVID-19 outbreak in
Indian states are tabulated in Table.\ref{tab:Estimated-undetected-},
for the first and the second phases of the pandemic, respectively.
Among the 20 Indian states, Delhi had the highest number of undetected
population during the first phase of the pandemic. Also, the ratio
of undetected to detected population was nearly $10$, the highest
among all the states. Our results for India as a whole and Indian
states show a good agreement with the already reported results. The
results obtained from the analysis are found to be consistent with
the MWSIR and other modified SEIR models \cite{comp-1,comp-4,comp-5,comp-6}.
In this regard, we would like to mention other studies, which have
already been carried out to estimate the undetected or missed COVID-19
infection cases as estimated in Refs.\cite{comp-2,undetect-1,undetect-2,undetect-3,undetect-4,undetect-5}.
An early COVID-19 pandemic study in Europe estimates the actual undetected
count size varies within $3.93-7.94$ times in different parts of
Europe \cite{uncount-1}. Another early pandemic study reported the
actual number of infections may have been 1.5 to 2.029 times more
than the actual reported count in the United States and 1.44 to 2.06
times more in Canada \cite{uncount-2}. A seroprevalence study of
COVID-19 infection in rural districts of south India reveals 7 numbers
of undetected for every RT-PCR confirmed case \cite{undetect-1}.
Chaubey and his colleagues estimated the real case of COVID-19 infections
to be 17 times higher in the first phase of the infection from a serosurvey
in India \cite{undetect-2}. Thus our undetected count for Indian
states agrees quite well with the other seroprevalence reports \cite{undetect-1,undetect-2}.
For the first phase of the pandemic, it is reported elsewhere that
about $10-50$ cases have gone missing for every detected case \cite{comp1}.
This report on South Indian slum areas indicates that the ratio of
detected to undetected cases was almost $1:195$. However, it is indeed
very difficult to estimate the exact ratio of active undetected to
active detected cases in the absence of reliable data. Our PINN model
estimates this ratio of active undetected to active detected population
$\sim4.58$ for the first phase of the pandemic in India. This ratio
is found to be smaller than the previously reported results.

The symptomatic unreported rate $\epsilon$ is found to be higher
in the state of Orissa in comparison to the other states. During the
second phase of the pandemic, we have observed that the ratio of undetected
to detected population is the highest for the state of Uttar Pradesh.
Surprisingly, peaks \emph{per million} active undetected and active
detected cases are found to be higher for the state of Kerala. From
the analysis, it is also evident that the symptomatic unreported rate
is higher in the state of Telangana for the second surge of the outbreak
in India. The $U/I$ ratios for these 20 states are shown graphically
in Fig.\ref{fig:The-active-ratio}.

\begin{table}[H]
\begin{centering}
\caption{\label{tab:Estimated-undetected-}Estimated undetected $(U)$ and
detected $(I)$ active cases for the first (top) and second (bottom)
phases of the pandemic using the PINN analysis. The rate of unreported
undetected population is also estimated by the PINN model. The maximum
value for each column is highlighted in \textbf{bold}.}
~\\
\begin{tabular}{|>{\centering}p{0.4cm}|c|c|c|c|c|c|}
\hline 
\multicolumn{1}{|>{\centering}p{0.4cm}|}{\textbf{\vl No.}} & \textbf{State} & \textbf{Abbrev.} & $U_{{\rm peak}}$ (per Million) & $I_{{\rm peak}}$ (per Million) & $\epsilon$ & $U_{{\rm peak}}/I_{{\rm peak}}$\tabularnewline
\hline 
\hline 
1 & Andhra Pradesh & AP & $0.0068$ & $0.0020$ & $0.0410$ & $3.4000$\tabularnewline
\hline 
2 & Assam & AS & $0.0037$ & $0.0010$ & $0.0132$ & $3.7000$\tabularnewline
\hline 
3 & Bihar & BR & $0.0007$ & $0.0003$ & $0.0238$ & $2.3333$\tabularnewline
\hline 
4 & Chhattisgarh & CT & $0.0050$ & $0.0013$ & $0.0150$ & $3.8462$\tabularnewline
\hline 
5 & Delhi & DL & \textbf{$\bm{0.0221}$} & $0.0023$ & $0.0444$ & \textbf{$\bm{9.6087}$}\tabularnewline
\hline 
6 & Gujarat & GJ & $0.0013$ & $0.0003$ & $0$ & $4.3333$\tabularnewline
\hline 
7 & Haryana & HR & $0.0036$ & $0.0008$ & $0.0208$ & $4.5000$\tabularnewline
\hline 
8 & Jharkhand & JH & $0.0019$ & $0.0004$ & 0.0249 & $4.7500$\tabularnewline
\hline 
9 & Karnataka & KA & $0.0107$ & $0.0018$ & $0.0443$ & $5.9444$\tabularnewline
\hline 
10 & Kerala & KL & $0.0165$ & \textbf{$\bm{0.0027}$} & $0.0369$ & $6.1111$\tabularnewline
\hline 
11 & Maharashtra & MH & $0.0082$ & $0.0024$ & $0$ & $3.4167$\tabularnewline
\hline 
12 & Madhya Pradesh & MP & $0.0012$ & $0.0003$ & $0.0346$ & $4.0000$\tabularnewline
\hline 
13 & Orissa & OR & $0.0032$ & $0.0007$ & \textbf{$\bm{0.0607}$} & $4.5714$\tabularnewline
\hline 
14 & Punjab & PB & $0.0034$ & $0.0008$ & $0.0429$ & $4.2500$\tabularnewline
\hline 
15 & Rajasthan & RJ & $0.0033$ & $0.0004$ & $0.0374$ & $8.2500$\tabularnewline
\hline 
16 & Telangana & TG & $0.0027$ & $0.0008$ & $0.0090$ & $3.3750$\tabularnewline
\hline 
17 & Tamil Nadu & TN & $0.0018$ & $0.0007$ & $0.0587$ & $2.5714$\tabularnewline
\hline 
18 & Uttar Pradesh & UP & $0.0028$ & $0.0003$ & $0$ & $9.3333$\tabularnewline
\hline 
19 & Uttarakhand & UT & $0.0041$ & $0.0011$ & $0.0300$ & $3.7273$\tabularnewline
\hline 
20 & West Bengal & WB & $0.0019$ & $0.0004$ & $0.0230$ & $4.7500$\tabularnewline
\hline 
\end{tabular}\\
~\\
~
\par\end{centering}
\centering{}%
\begin{tabular}{|c|c|c|c|c|c|c|}
\hline 
\multicolumn{1}{|>{\centering}p{0.4cm}|}{\textbf{\vl No.}} & \textbf{State} & \textbf{Abbrev.} & $U_{{\rm peak}}$ (per Million) & $I_{{\rm peak}}$ (per Million) & $\epsilon$ & $U_{{\rm peak}}/I_{{\rm peak}}$\tabularnewline
\hline 
\hline 
1 & Andhra Pradesh & AP & $0.0060$ & $0.0040$ & $0.0309$ & 1.5000\tabularnewline
\hline 
2 & Assam & AS & $0.0025$ & $0.0015$ & $0.0301$ & $1.6667$\tabularnewline
\hline 
3 & Bihar & BR & $0.0015$ & $0.0010$ & $0.0288$ & $1.5000$\tabularnewline
\hline 
4 & Chhattisgarh & CT & $0.0039$ & $0.0047$ & 0.0407 & $0.8298$\tabularnewline
\hline 
5 & Delhi & DL & $0.0093$ & $0.0055$ & $0.0407$ & $1.6909$\tabularnewline
\hline 
6 & Gujarat & GJ & $0.0071$ & $0.0025$ & $0.0438$ & $2.8400$\tabularnewline
\hline 
7 & Haryana & HR & $0.0041$ & $0.0041$ & $0.0229$ & $1.0000$\tabularnewline
\hline 
8 & Jharkhand & JH & $0.0031$ & $0.0017$ & $0.0289$ & $1.8235$\tabularnewline
\hline 
9 & Karnataka & KA & $0.0138$ & $0.0094$ & $0.0360$ & $1.4681$\tabularnewline
\hline 
10 & Kerala & KL & \textbf{$\bm{0.0159}$} & \textbf{$\bm{0.0128}$} & $0.0443$ & $1.2422$\tabularnewline
\hline 
11 & Maharashtra & MH & $0.0082$ & $0.0059$ & $0.0421$ & $1.3898$\tabularnewline
\hline 
12 & Madhya Pradesh & MP & $0.0026$ & $0.0013$ & $0.0480$ & $2.0000$\tabularnewline
\hline 
13 & Orissa & OR & $0.0039$ & $0.0022$ & $0.0390$ & $1.7727$\tabularnewline
\hline 
14 & Punjab & PB & $0.0037$ & $0.0027$ & $0.0532$ & $1.3704$\tabularnewline
\hline 
15 & Rajasthan & RJ & 0.0037 & $0.0027$ & $0.0317$ & $1.3704$\tabularnewline
\hline 
16 & Telangana & TG & $0.0045$ & $0.0022$ & \textbf{$\bm{0.0553}$} & $2.0455$\tabularnewline
\hline 
17 & Tamil Nadu & TN & $0.0054$ & $0.0040$ & $0.0436$ & 1.3500\tabularnewline
\hline 
18 & Uttar Pradesh & UP & $0.0041$ & $0.0014$ & $0.0457$ & \textbf{$\bm{2.9286}$}\tabularnewline
\hline 
19 & Uttarakhand & UT & $0.0071$ & $0.0075$ & $0.0330$ & $0.9467$\tabularnewline
\hline 
20 & West Bengal & WB & $0.0012$ & $0.0013$ & $0.0369$ & $0.9231$\tabularnewline
\hline 
\end{tabular}
\end{table}

But for some of the Indian states such as Uttar Pradesh and Delhi,
this ratio is found to be $\sim10$, which is comparable to the reported
results \cite{comp1}. For the second phase of the pandemic, the active
$U/I$ ratio is found to be $\sim1.3$ for India. The pandemic became
deadliest during the second phase of the pandemic and most of the
infectious population needed medical attention to cure the disease,
for which the detected infectious count will go higher, which has
caused the $U/I$ ratio to decrease, compared to that during the first
phase.

\begin{figure}[t]
\begin{centering}
\includegraphics[width=1\textwidth]{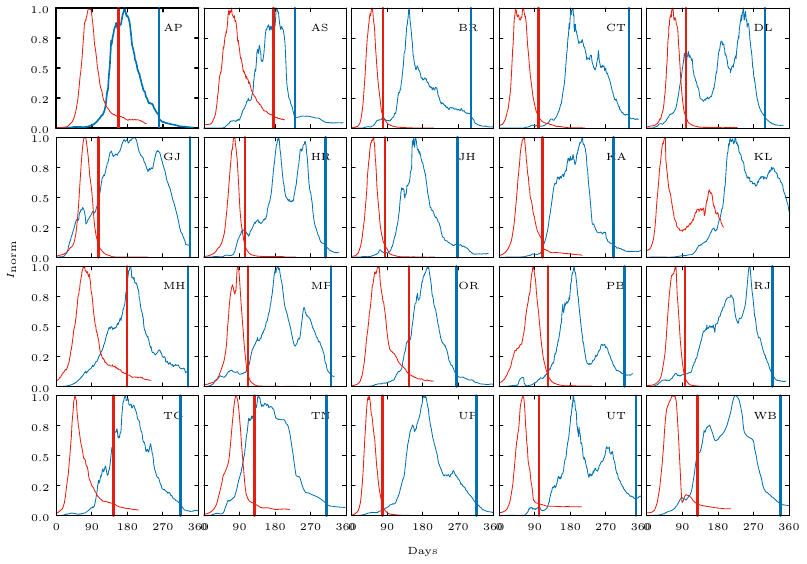}\vskip-20pt~
\par\end{centering}
\caption{\label{fig:The-dataset-used}Graphical representation of the dataset
used for ANN-based forecast along with the target values shown by
the vertical lines. The blue and red colors represent the first and
the second phases of the pandemic.}
\end{figure}

\section{4. ANN-based forecasting scenario}

We now test the predictability feature of the ANN-based model with
inputs from our PINN model. As mentioned before, our primary premise
regarding the two phases of the COVID-19 pandemic in India is that
the overall physical situations during both phases are similar except
for the more aggressive nature of infection during the second phase.
The evolution of the pandemic in terms of active infectives $(I)$
during these two phases in the $20$ worst affected states in India
is shown in Fig.\ref{fig:The-dataset-used}. The horizontal axes in
all the panels indicate the number of days starting from the zeroth
day of the pandemic in the respective phases. The \emph{zeroth} day
is the day at which we consider the pandemic to have started. The
first and second phases are respectively shown in blue and red colours.
The vertical axes are the number of active infective cases $(I_{{\rm norm}})$
normalized to unity in both phases. The normalization helps us reduce
both the datasets for the number of active infective cases to a single
framework for further processing through the neural network, which
is similar to standardization and pre-processing of the data. Altogether,
we have now $40$ sets of data which are to be processed through the
neural network. As can be seen from the figure, the shapes of the
curves for active cases are almost similar in both phases except for
the fact that the curves are wider for the first phase.

\begin{table}[t]
\caption{\label{tab:The-dataset-used}The dataset for phase \#1 (top) \& \#2
(bottom), used for forecasting using ANN.}
~\\

\centering{}%
\begin{tabular}{|c|c|c|c|c|c|c|r@{\extracolsep{0pt}.}l|c|}
\cline{2-10} \cline{3-10} \cline{4-10} \cline{5-10} \cline{6-10} \cline{7-10} \cline{8-10} \cline{10-10} 
\multicolumn{1}{c|}{} & \multirow{2}{*}{State} & \multirow{2}{*}{Target} & \multicolumn{7}{c|}{\vl Features}\tabularnewline
\cline{4-10} \cline{5-10} \cline{6-10} \cline{7-10} \cline{8-10} \cline{10-10} 
\multicolumn{1}{c|}{} &  &  & $D_{{\rm peak}}$ & \vl\makebox[1.5cm]{$I_{\rm peak}$} & \makebox[1.5cm]{$\beta_0$} & \makebox[1.5cm]{$\beta_{\rm peak}$} & \multicolumn{2}{c|}{\makebox[1.5cm]{$\epsilon$}} & \makebox[1.5cm]{\vl$\frac{U_{\rm peak}}{I_{\rm peak}}$}\tabularnewline
\hline 
\multirow{20}{*}{\makebox[0.8cm]{Phase I}} & AP & $260$ & $173$ & $103701$ & $0.00185$ & $0.00012$ & \multicolumn{2}{c|}{$0.0410$} & $3.4000$\tabularnewline
\cline{2-10} \cline{3-10} \cline{4-10} \cline{5-10} \cline{6-10} \cline{7-10} \cline{8-10} \cline{10-10} 
 & AS & $230$ & $183$ & $34393$ & $0.00226$ & $0.00019$ & \multicolumn{2}{c|}{$0.0132$} & $3.7000$\tabularnewline
\cline{2-10} \cline{3-10} \cline{4-10} \cline{5-10} \cline{6-10} \cline{7-10} \cline{8-10} \cline{10-10} 
 & BR & $303$ & $146$ & $32716$ & $0.00082$ & $0.00007$ & \multicolumn{2}{c|}{$0.0238$} & $2.3333$\tabularnewline
\cline{2-10} \cline{3-10} \cline{4-10} \cline{5-10} \cline{6-10} \cline{7-10} \cline{8-10} \cline{10-10} 
 & CT & $329$ & $187$ & $38198$ & $0.00237$ & $0.00022$ & \multicolumn{2}{c|}{$0.0150$} & $3.8462$\tabularnewline
\cline{2-10} \cline{3-10} \cline{4-10} \cline{5-10} \cline{6-10} \cline{7-10} \cline{8-10} \cline{10-10} 
 & DL & $299$ & $245$ & $44456$ & $0.00329$ & $0.00022$ & \multicolumn{2}{c|}{$0.0444$} & $9.6087$\tabularnewline
\cline{2-10} \cline{3-10} \cline{4-10} \cline{5-10} \cline{6-10} \cline{7-10} \cline{8-10} \cline{10-10} 
 & GJ & $338$ & $199$ & $16735$ & $0.00129$ & $0.00009$ & \multicolumn{2}{c|}{$0$} & $4.3333$\tabularnewline
\cline{2-10} \cline{3-10} \cline{4-10} \cline{5-10} \cline{6-10} \cline{7-10} \cline{8-10} \cline{10-10} 
 & HR & $308$ & $189$ & $21683$ & $0.00191$ & $0.00024$ & \multicolumn{2}{c|}{$0.0208$} & $4.5000$\tabularnewline
\cline{2-10} \cline{3-10} \cline{4-10} \cline{5-10} \cline{6-10} \cline{7-10} \cline{8-10} \cline{10-10} 
 & JH & $268$ & $162$ & $15701$ & $0.00251$ & $0.00019$ & \multicolumn{2}{c|}{$0.0249$} & $4.7500$\tabularnewline
\cline{2-10} \cline{3-10} \cline{4-10} \cline{5-10} \cline{6-10} \cline{7-10} \cline{8-10} \cline{10-10} 
 & KA & $290$ & $210$ & $120940$ & $0.00106$ & $0.00008$ & \multicolumn{2}{c|}{$0.0443$} & $5.9444$\tabularnewline
\cline{2-10} \cline{3-10} \cline{4-10} \cline{5-10} \cline{6-10} \cline{7-10} \cline{8-10} \cline{10-10} 
 & KL & $380$ & $224$ & $97525$ & $0.00170$ & $0.00014$ & \multicolumn{2}{c|}{$0.0369$} & $6.1111$\tabularnewline
\cline{2-10} \cline{3-10} \cline{4-10} \cline{5-10} \cline{6-10} \cline{7-10} \cline{8-10} \cline{10-10} 
 & MH & $334$ & $187$ & $302136$ & $0.00061$ & $0.00005$ & \multicolumn{2}{c|}{$0$} & $3.4167$\tabularnewline
\cline{2-10} \cline{3-10} \cline{4-10} \cline{5-10} \cline{6-10} \cline{7-10} \cline{8-10} \cline{10-10} 
 & MP & $322$ & $187$ & $22811$ & $0.00077$ & $0.00008$ & \multicolumn{2}{c|}{$0.0346$} & $4.0000$\tabularnewline
\cline{2-10} \cline{3-10} \cline{4-10} \cline{5-10} \cline{6-10} \cline{7-10} \cline{8-10} \cline{10-10} 
 & OR & $266$ & $192$ & $35039$ & $0.00213$ & $0.00013$ & \multicolumn{2}{c|}{$0.0607$} & $4.5714$\tabularnewline
\cline{2-10} \cline{3-10} \cline{4-10} \cline{5-10} \cline{6-10} \cline{7-10} \cline{8-10} \cline{10-10} 
 & PB & $317$ & $189$ & $22399$ & $0.00215$ & $0.00021$ & \multicolumn{2}{c|}{$0.0429$} & $4.2500$\tabularnewline
\cline{2-10} \cline{3-10} \cline{4-10} \cline{5-10} \cline{6-10} \cline{7-10} \cline{8-10} \cline{10-10} 
 & RJ & $318$ & $260$ & $28758$ & $0.00071$ & $0.00004$ & \multicolumn{2}{c|}{$0.0374$} & $8.2500$\tabularnewline
\cline{2-10} \cline{3-10} \cline{4-10} \cline{5-10} \cline{6-10} \cline{7-10} \cline{8-10} \cline{10-10} 
 & TG & $315$ & $174$ & $32484$ & $0.00207$ & $0.00019$ & \multicolumn{2}{c|}{$0.0090$} & $3.3750$\tabularnewline
\cline{2-10} \cline{3-10} \cline{4-10} \cline{5-10} \cline{6-10} \cline{7-10} \cline{8-10} \cline{10-10} 
 & TN & $310$ & $139$ & $57968$ & $0.00115$ & $0.00013$ & \multicolumn{2}{c|}{$0.0587$} & $2.5714$\tabularnewline
\cline{2-10} \cline{3-10} \cline{4-10} \cline{5-10} \cline{6-10} \cline{7-10} \cline{8-10} \cline{10-10} 
 & UP & $317$ & $187$ & $68235$ & $0.00046$ & $0.00002$ & \multicolumn{2}{c|}{$0$} & $9.3333$\tabularnewline
\cline{2-10} \cline{3-10} \cline{4-10} \cline{5-10} \cline{6-10} \cline{7-10} \cline{8-10} \cline{10-10} 
 & UT & $346$ & $188$ & $12644$ & $0.00586$ & $0.00058$ & \multicolumn{2}{c|}{$0.0300$} & $3.7273$\tabularnewline
\cline{2-10} \cline{3-10} \cline{4-10} \cline{5-10} \cline{6-10} \cline{7-10} \cline{8-10} \cline{10-10} 
 & WB & $338$ & $223$ & $37190$ & $0.00062$ & $0.00005$ & \multicolumn{2}{c|}{$0.0230$} & $4.7500$\tabularnewline
\hline 
\end{tabular}\\
~\\
~\\
\begin{tabular}{|c|c|c|c|c|c|c|r@{\extracolsep{0pt}.}l|c|}
\cline{2-10} \cline{3-10} \cline{4-10} \cline{5-10} \cline{6-10} \cline{7-10} \cline{8-10} \cline{10-10} 
\multicolumn{1}{c|}{} & \multirow{2}{*}{State} & \multirow{2}{*}{Target} & \multicolumn{7}{c|}{\vl Features}\tabularnewline
\cline{4-10} \cline{5-10} \cline{6-10} \cline{7-10} \cline{8-10} \cline{10-10} 
\multicolumn{1}{c|}{} &  &  & $D_{{\rm peak}}$ & \vl\makebox[1.5cm]{$I_{\rm peak}$} & \makebox[1.5cm]{$\beta_0$} & \makebox[1.5cm]{$\beta_{\rm peak}$} & \multicolumn{2}{c|}{\makebox[1.5cm]{$\epsilon$}} & \makebox[1.5cm]{\vl$\frac{U_{\rm peak}}{I_{\rm peak}}$}\tabularnewline
\hline 
\multirow{20}{*}{\makebox[0.8cm]{Phase II}} & AP & $158$ & $82$ & $211554$ & $0.00228$ & $0.00024$ & \multicolumn{2}{c|}{$0.0309$} & $1.5000$\tabularnewline
\cline{2-10} \cline{3-10} \cline{4-10} \cline{5-10} \cline{6-10} \cline{7-10} \cline{8-10} \cline{10-10} 
 & AS & $176$ & $70$ & $56188$ & $0.00467$ & $0.00035$ & \multicolumn{2}{c|}{$0.0301$} & $1.6667$\tabularnewline
\cline{2-10} \cline{3-10} \cline{4-10} \cline{5-10} \cline{6-10} \cline{7-10} \cline{8-10} \cline{10-10} 
 & BR & $79$ & $51$ & $115152$ & $0.00167$ & $0.00012$ & \multicolumn{2}{c|}{$0.0288$} & $1.5000$\tabularnewline
\cline{2-10} \cline{3-10} \cline{4-10} \cline{5-10} \cline{6-10} \cline{7-10} \cline{8-10} \cline{10-10} 
 & CT & $99$ & $60$ & $131245$ & $0.00323$ & $0.00037$ & \multicolumn{2}{c|}{$0.0407$} & $0.8298$\tabularnewline
\cline{2-10} \cline{3-10} \cline{4-10} \cline{5-10} \cline{6-10} \cline{7-10} \cline{8-10} \cline{10-10} 
 & DL & $99$ & $64$ & $99752$ & $0.00739$ & $0.00073$ & \multicolumn{2}{c|}{$0.0407$} & $1.6909$\tabularnewline
\cline{2-10} \cline{3-10} \cline{4-10} \cline{5-10} \cline{6-10} \cline{7-10} \cline{8-10} \cline{10-10} 
 & GJ & $107$ & $72$ & $148296$ & $0.00189$ & $0.00027$ & \multicolumn{2}{c|}{$0.0438$} & $2.8400$\tabularnewline
\cline{2-10} \cline{3-10} \cline{4-10} \cline{5-10} \cline{6-10} \cline{7-10} \cline{8-10} \cline{10-10} 
 & HR & $104$ & $78$ & $116868$ & $0.00307$ & $0.00043$ & \multicolumn{2}{c|}{$0.0229$} & $1.0000$\tabularnewline
\cline{2-10} \cline{3-10} \cline{4-10} \cline{5-10} \cline{6-10} \cline{7-10} \cline{8-10} \cline{10-10} 
 & JH & $85$ & $56$ & $61168$ & $0.00562$ & $0.00030$ & \multicolumn{2}{c|}{$0.0289$} & $1.8235$\tabularnewline
\cline{2-10} \cline{3-10} \cline{4-10} \cline{5-10} \cline{6-10} \cline{7-10} \cline{8-10} \cline{10-10} 
 & KA & $110$ & $62$ & $605507$ & $0.00192$ & $0.00021$ & \multicolumn{2}{c|}{$0.0360$} & $1.4681$\tabularnewline
\cline{2-10} \cline{3-10} \cline{4-10} \cline{5-10} \cline{6-10} \cline{7-10} \cline{8-10} \cline{10-10} 
 & KL & $-$ & $46$ & $445697$ & $0.00349$ & $0.00044$ & \multicolumn{2}{c|}{$0.0443$} & $1.2422$\tabularnewline
\cline{2-10} \cline{3-10} \cline{4-10} \cline{5-10} \cline{6-10} \cline{7-10} \cline{8-10} \cline{10-10} 
 & MH & $179$ & $69$ & $701615$ & $0.00063$ & $0.00012$ & \multicolumn{2}{c|}{$0.0421$} & $1.3898$\tabularnewline
\cline{2-10} \cline{3-10} \cline{4-10} \cline{5-10} \cline{6-10} \cline{7-10} \cline{8-10} \cline{10-10} 
 & MP & $112$ & $86$ & $111365$ & $0.00122$ & $0.00013$ & \multicolumn{2}{c|}{$0.0480$} & $2.0000$\tabularnewline
\cline{2-10} \cline{3-10} \cline{4-10} \cline{5-10} \cline{6-10} \cline{7-10} \cline{8-10} \cline{10-10} 
 & OR & $145$ & $67$ & $100235$ & $0.00316$ & $0.00029$ & \multicolumn{2}{c|}{$0.0390$} & $1.7727$\tabularnewline
\cline{2-10} \cline{3-10} \cline{4-10} \cline{5-10} \cline{6-10} \cline{7-10} \cline{8-10} \cline{10-10} 
 & PB & $124$ & $85$ & $79963$ & $0.00226$ & $0.00042$ & \multicolumn{2}{c|}{$0.0532$} & $1.3704$\tabularnewline
\cline{2-10} \cline{3-10} \cline{4-10} \cline{5-10} \cline{6-10} \cline{7-10} \cline{8-10} \cline{10-10} 
 & RJ & $96$ & $73$ & $212753$ & $0.00170$ & $0.00014$ & \multicolumn{2}{c|}{$0.0317$} & $1.3704$\tabularnewline
\cline{2-10} \cline{3-10} \cline{4-10} \cline{5-10} \cline{6-10} \cline{7-10} \cline{8-10} \cline{10-10} 
 & TG & $145$ & $47$ & $80185$ & $0.00510$ & $0.00035$ & \multicolumn{2}{c|}{$0.0553$} & $2.0455$\tabularnewline
\cline{2-10} \cline{3-10} \cline{4-10} \cline{5-10} \cline{6-10} \cline{7-10} \cline{8-10} \cline{10-10} 
 & TN & $128$ & $81$ & $313048$ & $0.00110$ & $0.00017$ & \multicolumn{2}{c|}{$0.0436$} & $1.3500$\tabularnewline
\cline{2-10} \cline{3-10} \cline{4-10} \cline{5-10} \cline{6-10} \cline{7-10} \cline{8-10} \cline{10-10} 
 & UP & $78$ & $45$ & $310783$ & $0.00120$ & $0.00006$ & \multicolumn{2}{c|}{$0.0457$} & $2.9286$\tabularnewline
\cline{2-10} \cline{3-10} \cline{4-10} \cline{5-10} \cline{6-10} \cline{7-10} \cline{8-10} \cline{10-10} 
 & UT & $101$ & $60$ & $85120$ & $0.01085$ & $0.00118$ & \multicolumn{2}{c|}{$0.0330$} & $0.9467$\tabularnewline
\cline{2-10} \cline{3-10} \cline{4-10} \cline{5-10} \cline{6-10} \cline{7-10} \cline{8-10} \cline{10-10} 
 & WB & $128$ & $70$ & $132181$ & $0.00088$ & $0.00011$ & \multicolumn{2}{c|}{$0.0369$} & $0.9231$\tabularnewline
\hline 
\end{tabular}
\end{table}

\begin{figure}[t]
\begin{centering}
\includegraphics[width=0.5\textwidth]{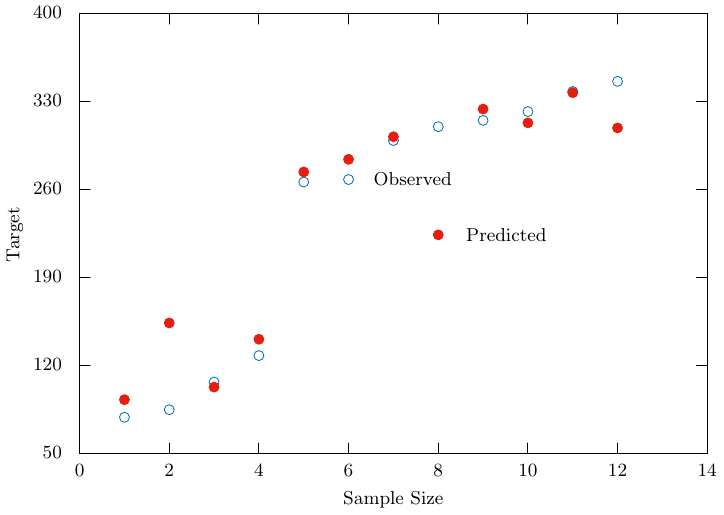}\hfill{}\includegraphics[width=0.5\textwidth]{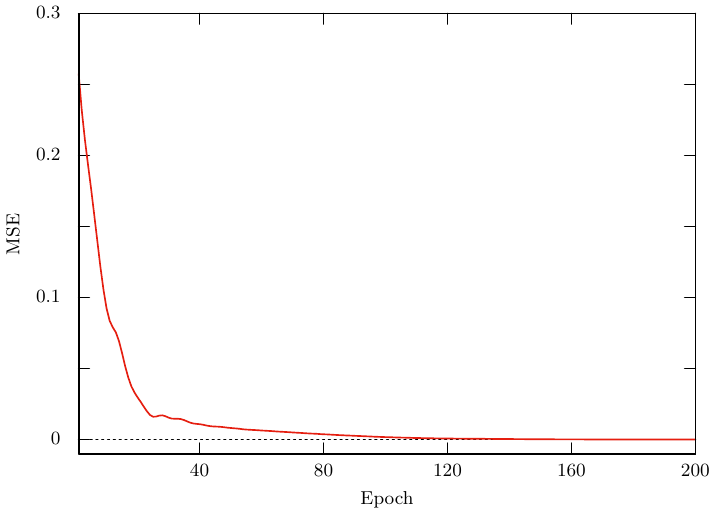}
\par\end{centering}
\caption{\label{fig:The-ANN-predictions}The ANN predictions for various sample
sizes (left) and the MSE of the optimization (right). The samples
are here the target values of the twenty states (see text). Except
for two values, the rest of the predictions quite agree with the observed
values.}
\end{figure}

The most common and logical information to be predicted during such
an evolving pandemic is to have an idea of the timeline for when the
pandemic is going to subside or end. Toward this, we have constructed
the \emph{target} values for our \emph{prediction} as the day when
the number of active infections reduces to $10\%$ of the daily active
peak value for any particular state. These \emph{target} values are
shown graphically in Fig.\ref{fig:The-dataset-used}, by the vertical
lines. We have identified six \emph{features} for the ANN model, which
inputs are, peak day (the day at which the active cases reach their
peak), the number of cases recorded at the peak day, $\beta_{t}$
at the zeroth day, $\beta_{t}$ at the peak day, $\epsilon$(symptomatic
unreported rate) and the ratio of the undetected active cases to the
detected active cases $(U_{{\rm peak}}/I_{{\rm peak}})$. The transmission
rates $\beta_{0,t}$ are calculated numerically from our new SEIUR
COVID-19 model. The $(U_{{\rm peak}}/I_{{\rm peak}})$ ratio and $\epsilon$
values are taken from the PINN results. Except for the peak day and
active cases at the peak day, all the inputs used for forecasting
are unique as they are obtained from our new model and PINN analysis.
We combine both the first phase data and second phase data of the
COVID-19 outbreak for the Indian states and the dataset is supplied
to the ANN analysis. We have used $70\%$ of the data to train the
model and the rest of the $30\%$ is used for prediction purposes.
We have used $6$ hidden layers with $64$ neurons each and the epoch
used is $200.$ The loss function used is mean squared loss (MSE)
and optimization is carried out with the help of the Adam optimizer.
The rectilinear unit (ReLU) is used as an activation function. The
dataset used for forecasting is tabulated in Table.\ref{tab:The-dataset-used}.
The prediction results are shown in the left panel of Fig.\ref{fig:The-ANN-predictions}.
The root mean squared error (RMSE) for the analysis is found to be
$\sim11.5\%$. The results of the forecast scenario are shown in Fig.\ref{fig:The-ANN-predictions}.

\section{5. Conclusion}

In this work, we have made an attempt to estimate the infectious population
of the COVID-19 outbreak in India during the years 2020 and 2021,
who goes undetected and do not contribute to the publicly and officially
available records. To this effect, we have constructed an new SEIUR
model by adding two new compartments `undetected' and `undetected
removed' to our previously developed and successfully deployed model
\cite{OurModel}. One of the novelties of this work is the use of
a new tool, PINN \cite{lagaris}, employed to solve the SEIUR model
and estimate the undetected population through `parameter discovery'.
The estimation of the undetected infectious population itself is considered
as an important finding with reference to such kinds of pandemics
in India, which is now the most populous country. We performed the
PINN analysis for 20 worst affected Indian states and estimated the
undetected population. The ratio of active undetected $(U)$ to the
active detected $(I)$ cases are calculated for the states as well
as for India as a whole. The $(U/I)$ ratio is highest in the state
of Delhi $(\sim10)$ and the state of Uttar Pradesh $(\sim3)$ in
the first phase and the second phase, respectively. For India as a
whole, this ratio is $4.58$ and $1.3$ for the first and second waves
of the pandemic, respectively.

One important finding of this PINN analysis is the estimation of the
rate at which symptomatic infectious population goes undetected but
contribute equally to the spread of the disease. This rate $\epsilon$
is found to be the highest in the state of Orissa $(0.0607)$ and
the state of Telangana $(0.0553)$ in the first and the second phases,
respectively. The value of $\epsilon$ for India is found to be $0.0484$
and $0.0493$ for the first and the second phases, respectively. We
can say that this $\epsilon$ parameter is a measure of the \emph{law-abiding
disciplinary index} of the country as well as for Indian states. The
active $(U/I)$ ratio and the $\epsilon$ gives a clear picture of
the response of the Indian states and India as a whole towards the
tackling of the future COVID-19 or similar outbreaks.

\subsection{{\bf 5.1 Strength and weakness of the approach}}

We note that any modelling of disease outbreaks invariably has to
employ some kind of epidemiological model \cite{sir,seir1,seir2}.
However, depending on the complexity of an outbreak, especially when
it is relatively newer like the recent COVID-19 pandemic, the model
parameters may vary widely. Also, the more the number of independent
parameters of an outbreak, the more involved is the dynamical model.
In the economically developed countries with well-defined healthcare
systems, the outbreak data are usually reliable and detailed \cite{italy-1,italy-2,italy-3}.
In contrast to this, for countries with developing economies and relatively
unorganised healthcare systems, fine-scale accurate data such as health-related
data of hospitalised patients, number of patients requiring various
levels of intensive care etc. are extremely difficult to obtain. In
such cases, the epidemiological models have to be based on certain
loosely defined parameters and one looks forward to replicating the
available outbreak data over a broad timeline. It is where the ANN-based
methods such as the PINN analysis find their places, where one has
to determine a variety of parameters with high degrees of uncertainty.
ANN-based methods have the capacity of modelling such data with minimal
effort. And we believe, we have shown a relatively novel way of such
`parameter discovery' with the PINN-based epidemiological SEIUR model.

Naturally, the PINN-based method requires large amount of outbreak
data to have a successful forecasting. With increasing awareness and
preparedness, many countries now have a lot of post-COVID-19 measures,
which will be able to yield reliable data for future outbreaks. As
number of available datasets increases, the accuracy of ANN-based
modelling also increases.

\section*{Declaration}

The authors have no conflicts to disclose.

\section*{Data availability}

All data used in this work are publicly available. Specific data sets
will be made available on request.

\section*{Acknowledgement}

This work is supported by SERB (DST, India) research project grant
CRG/2018/002971.

%\bibliographystyle{spmpsci}
%\bibliography{references}

\end{document}